\documentclass[12pt]{article}

\def\be{\begin{equation}}
\def\ee{\end{equation}}
\def\bea{\begin{eqnarray}}
\def\eea{\end{eqnarray}}
\def\vk{{\bf k}}
\def\vko{{\bf k}_{1}}
\def\vkt{{\bf k}_{2}}

\begin{document}

\begin{titlepage}

\begin{centering}

\vfill

{\bf REACTION OF THE FERMION FIELD \\ON SPONTANEOUS CHANGE OF ITS MASS }\\

\vspace{1cm}

 I.V.ANDREEV\\
\vspace{0.5cm}

{\it
P.N.Lebedev Physical Institute, Moscow, Russia}\\
\vspace{3cm}
\centerline{\bf Abstract}
\end{centering}
\vspace{0.3cm}\noindent

Modification of the particles
in the course of the source evolution is considered. Influence of this
effect on multiplicities and correlations of the particles is displayed,
including an enhancement of the production rates and identical particle
 correlations
and also back-to-back particle-antiparticle correlations.

\vfill \vfill

\end{titlepage}

\section{Introduction}

It is well known that in quantum chromodynamics the chiral invariance is
spontaneously broken at low energies and the fermion masses (say nucleon or
constituent quark mass) arise essentially due to chiral simmetry breaking.
In the course of phase transition at high fermion density or at high
temperature the approximate chiral invariance is restored and the light
$(u,d,s)$ quark masses become small (ensuring small explicit violation of
the symmetry). In practice the change of the fermion masses is treated 
(semi)classically.
 However it appears that, if the change of the mass is
fast enough, then the quantum fermion field is changing in a specific way
producing correlated fermion-antifermion pairs.

 Similar effects were considered earlier for bosons (mesons~\cite{AW,AC,A1,ACG}
 and photons~\cite{A2}) with
application to heavy-ion collisions~\cite{A2,A3}.  
 Below we consider 
the case of spatially homogeneous large volume. Applications
of this effect to heavy ion collisions will be given elsewhere.

Decomposition of the free fermion field is taken in the form:
\be
\psi(x)=\int\frac{d^{3}k}{(2\pi)^{3/2}}e^{i\vk{\bf x}}\sum_{\nu=1}^{2}[
u_{\nu}({\vk})b_{\nu}({\vk})e^{-iEt}+v_{\nu}(-{\vk})d^{\dag}_{\nu}(-{\vk})e^{iEt}]
\label{eq:1}
\ee             
where $b_{\nu},b^{\dag}_{\nu}$ and $d_{\nu},d^{\dag}_{\nu}$ are annihilation
and creation operators of particles and antiparticles, obeying standard
anticommutation relations.
%\be
%[b_{\nu}({\vko}),b^{\dag}_{\mu}({\vkt})]_{+}=\delta_{\mu\nu}\delta({\vko}-{\vkt}), \quad        
%[d_{\nu}({\vko}),d^{\dag}_{\mu}({\vkt})]_{+}=\delta_{\mu\nu}\delta({\vko}-{\vkt})
%\label{eq:2}
%\ee                                              
%Index $\nu=1,2$ corresponds to definite spin projection on the third axis of
%the coordinate system which is equal to $S_{3}=+1/2,-1/2$ for $b,b^{\dag}$-operators
%and $S_{3}=-1/2,+1/2$ for $d,d^{\dag}$-operators.
% and we choose the third coordinate axis along the fermion momentum ${\vk}$.

Bispinors $u_{\nu}({\vk}), v_{\mu}({\vk})$ are orthonormal,
%\be
%u^{\dag}_{\nu}({\vk})u_{\mu}({\vk})=v^{\dag}_{\nu}({\vk})v_{\mu}({\vk})=\delta_{\mu\nu},
%\label{eq:3}
%\ee
and bispinors $u_{\nu}({\vk})$ and $v_{\mu}({-\vk})$ having opposite momenta 
are orthogonal.
%\be
%u^{\dag}_{\nu}({\vk})v_{\mu}({-\vk})=v^{\dag}_{\nu}({\vk})u_{\mu}({-\vk})=0
%\label{eq:4}
%\ee
For these bispinors related to Dirac equation
\be
\left(i\gamma^{0}\frac{\partial}{\partial t}-\gamma^{n}k^{n}-m\right)\psi(\vk,t)=0,
\quad n=1,2,3
\label{eq:2}
\ee
we take the standard representation
\bea
u_{\nu}({\vk})=\frac{1}{N}(m+\gamma k)\left( {s_{\nu}}\atop 0\right),\quad 
v_{\nu}({\vk})=\frac{1}{N}(m-\gamma k)\left( 0\atop {s_{\nu}}\right)
\label{eq:3}
\eea
with two-component unit spinors $(s_{1})_{\rho}=\delta_{1\rho},\;
(s_{2})_{\rho}=\delta_{2\rho}$ and normalization factor
\be
N\equiv N(E,m)=\left( 2E(E+m)\right)^{1/2}
\label{eq:4}
\ee

\section{Step-like variation of the fermion mass}

Let the mass $m$ in (5) depends on time, $m=m(t)$.
We consider in this section the step-like variation of
the mass. This simple case reveals the main features of the phenomenon.
Let at time $t=0$ the mass changes instantly from $m_{i}$ to $m_{f}$.
According to Dirac equation (2),
the time derivative of the field $\psi$ has jump discontinuity at $t=0$.
Therefore the field $\psi(\vk,t)$ is continuous at the point $t=0$ ,
\be
\psi_{i}(\vk,-\delta t)=\psi_{f}(\vk,+\delta t),\quad \delta t\rightarrow~0
\label{eq:5}
\ee
Turning to decomposition (1) we choose the third coordinate axis along the
direction of momentum ${\vk}$:
\be
k^{(3)}=k=|{\vk}|
\label{eq:6}
\ee
and use for $u_{\nu}, v_{\nu}$ the representation (3). In this case the
bispinors $u_{\nu}, v_{\nu}$ correspond to definite helicities $h$
($h=1/2$ for $\nu=1$ and $h=-1/2$ for $\nu=2$) and Eq.(1) gives decomposition
over states with definite momentum and helicity. In view of continuity of
the field $\psi$ we have
\be
u_{i,\nu}(\vk)b_{i,\nu}(\vk)+v_{i,\nu}(-\vk)d^{\dag}_{i,\nu}(-\vk)=
u_{f,\nu}(\vk)b_{f,\nu}(\vk)+v_{f,\nu}(-\vk)d^{\dag}_{f,\nu}(-\vk)
\label{eq:7}
\ee
(sum over $\nu$).
Multiplying both sides of (7) by $u^{\dag}_{f,\mu}(\vk)$ and then by
$v^{\dag}_{f,\mu}(-\vk)$ and using orthogonality conditions we get the Bogoliubov transformation
of the annihilation and creation operators expressing final-state operators
through initial-state operators:
\bea
b_{f,\nu}(\vk)=\alpha_{\nu}(\vk)b_{i,\nu}(\vk)+\beta_{\nu}(\vk)d^{\dag}_{i,\nu}(-\vk), \nonumber \\
d^{\dag}_{f,\nu}(-\vk)=-\beta_{\nu}(\vk)b_{i,\nu}(\vk)+\alpha_{\nu}(\vk)d_{i,\nu}(-\vk)
\label{eq:8}
\eea
for every $\nu,{\vk}$, with coefficients
\bea
\alpha_{\nu}(\vk)=u^{\dag}_{f,\nu}(\vk)u_{i,\nu}(\vk)=v^{\dag}_{f,\nu}(-\vk)v_{i,\nu}(-\vk)
=\frac{(E_{f}+m_{f})(E_{i}+m_{i})+k^{2}}{N_{i}N_{f}}, \nonumber \\
\beta_{\nu}(\vk)=u^{\dag}_{f,\nu}(\vk)v_{i,\nu}(-\vk)=-v^{\dag}_{f,\nu}(-\vk)u_{i,\nu}(\vk)
=\mp\frac{k(E_{f}+m_{f}-E_{i}-m_{i})}{N_{i}N_{f}}
\label{eq:9}
\eea
for $\nu=1,2$ correspondingly where $N_{i},N_{f}$ are normalization factors of the
spinors $u_{\nu},v_{\nu}$ in the initial and final states, given by (4).
%\be
%N_{i}^{2}=2E_{i}(E_{i}+m_{i}),\quad N_{f}^{2}=2E_{f}(E_{f}+m_{f})
%\label{eq:13}
%\ee
As can be seen from (9), the Bogoliubov coefficients $\alpha_{\nu}(\vk),
\beta_{\nu}(\vk)$ satisfy the condition:
\be
|\alpha_{\nu}(\vk)|^{2}+|\beta_{\nu}(\vk)|^{2}=1
\label{eq:10}
\ee
so that the transformation (8) is unitary SU(2)-transformation, as it must be
to preserve the anticommutation relations.
% Let us note in this connection
%that, for standard choice (6) of bispinors,  
%the condition (14) is fulfilled for coefficients $\alpha_{\nu}(\vk),
%\beta_{\nu}(\vk)$, given by (12), only if (9) is valid.
%This feature confirms that
In the case (6)
the Bogoliubov transformation connects creation and annihilation
operators which have opposite directions of momenta and opposite spin
projections (equal helicities). It can be shown that the last statement
remains valid for any direction of the momentum $\vk$.
Let us note that for step-like transition
the Bogoliubov coefficients (9) are real-valued. The coefficient $\beta_{\nu}(\vk)$
changes its sign if we change the sign of the momentum ($\vk\to-\vk$), or change
$\nu (1\leftrightarrow 2)$, or interchange the initial and the final
states $(f\leftrightarrow i)$.

In general the coefficients $\alpha_{\nu}(\vk), \beta_{\nu}(\vk)$ of 
$SU(2)$-transformation (8) can be represented in the form
\be
\alpha(\vk)=\cos r(\vk)e^{i\varphi_{\alpha}},\qquad
\beta(\vk)=\sin r(\vk)e^{i\varphi_{\beta}}
\label{eq:11}
\ee
where
\be
r(\vk)=tan^{-1}|\beta/\alpha|
\label{eq:12}
\ee
is the main evolution parameter and the phases $\varphi_{\alpha},\varphi_{\beta}$
do not play important role and they will not be considered here. For step-like
transition the phases are absent.  

Using (6),(9) we get
\be
\alpha^{2}(\vk)=\frac{1}{2}+\frac{k^{2}+m_{i}m_{f}}{2E_{i}E_{f}}, \qquad
\beta^{2}(\vk)=\frac{1}{2}-\frac{k^{2}+m_{i}m_{f}}{2E_{i}E_{f}},
\label{eq:13}
\ee
As one can see from (13) the evolution parameter $r(\vk)$ is equal zero
at $\vk=0$, it is maximal at
\be
k^{2}=k_{m}^{2}=m_{i}m_{f}+\frac{m_{i}^{2}m_{f}^{2}}{(m_{i}-m_{f})^{2}}
\label{eq:14}
\ee
and it falls down slowly at large $k$,
\be
r(\vk)\approx \frac{|m_{i}-m_{f}|}{2k}
\label{eq:15}
\ee
In the point of the maximum the parameter $r$ depends only on the ratio
$m_{f}/m_{i}$ and it is sizable if the mass ratio is small (or large),
say $m_{i}\gg m_{f}$. In this case
\be
\tan r=|\beta/\alpha|\approx 1-2\sqrt{\frac{m_{f}}{m_{i}}} \quad {\rm for} \quad k^{2}\approx m_{i}m_{f}
\label{eq:16}
\ee
If one takes the fermion masses $m_{i}$ and $m_{f}$ to be the constituent quark
mass($\sim 350$ MeV) and the current quark mass ($\sim 5$ MeV) then
$\tan r_{max}\approx 0.76$ at $k\approx 42$ MeV.

In reality the change of the fermion mass has finite time duration $\tau$
and the step-like approximation is valid only if the momentum is much less than
inverse time duration, $k\tau\ll 1$. The effect of finite time duration
will be discussed in the next section.                                                                                                                                                                                                    

\section{Smooth transition}
Let us consider the smooth variation of the fermion mass $m(t)$. In this case 
the coefficients $\alpha(\vk),\beta(\vk)$ of the Bogoliubov transformation
can be expressed through solutions of the Dirac equation (2)~\cite{N}.
To solve the equation it is helpful to use its squared form, representing the
Dirac $\psi$-function in (5) in the form
\be
\psi(\vk,t)=\left( i\gamma^{0}\frac{d}{dt}-\vec{\gamma}\vec{k}+m(t)\right)
\chi(\vk,t)
\label{eq:17}
\ee
Then the equation (2) takes the form:
\be
\left(\frac{d^{2}}{dt^{2}}-i\gamma^{0}\frac{dm}{dt}+k^{2}+m^{2}(t)\right)
\chi(\vk,t)=0
\label{eq:18}
\ee
which splits into two complex-conjugated equations for two upper and two lower
components of $\chi$. The bispinor $\chi$ can be written in the form
\be
\chi_{\nu}(\vk,t)=\left( s_{\nu}\atop 0\right)\varphi(\vk,t)+
            \left( 0\atop s_{\nu}\right)\varphi(\vk,t),\; \nu=1,2
\label{eq:19}
\ee
where $s_{\nu}$ are unit two-component spinors, see (3). As a result we have
the second-order equation for scalar function $\varphi(\vk,t)$:
\be
\left(\frac{d^{2}}{dt^{2}}-i\frac{dm}{dt}+k^{2}+m^{2}(t)\right)
\varphi(\vk,t)=0
\label{eq:20}
\ee
which is nothing but oscillator equation with complex-valued variable 
frequency (energy).

Considering the smooth variation of the mass $m(t)$ we use the reference
model with
\be
m(t)=\frac{m_{f}+m_{i}}{2}+\left(\frac{m_{f}-m_{i}}{2}\right)tanh(2t/\tau)
\label{eq:21}
\ee
for which (20) has the exact solution and contains the important parameter $\tau$
giving characteristic time of the mass variation.
 The resulting solution (17) of the
classical Dirac equation (2) describes the evolution of the normalized fermion
field~\cite{N}. Choosing one of the upper spinors $s_{\nu}$ (say $s_{1}$)
we get 
\bea
\psi(\vk,t)=u_{1}(\vk,m_{i},E_{i})e^{-iE_{i}t} ,\; t<0 \nonumber \\
\psi(\vk,t)=\alpha_{1}(\vk)u_{1}(\vk,m_{f},E_{f})e^{-iE_{f}t}+
            \beta_{1}(\vk)u_{1}(\vk,m_{f},-E_{f})e^{iE_{f}t}, \; t>0
\label{eq:22}
\eea
showing appearance of the negative-energy wave (creation of antiparticles)
with correct Boboliubov coefficient~\cite{N}. The choice of another unit spinor  
in (22) gives similar result. The momentum $\vk$ in (22) may have arbitrary
direction. If we take the $\vk$-direction along the third (spin quantization)
axis then bispinors $u_{1}$ in (22) have definite helicity $h=1/2$ which is
conserved in the course of transition.
For the model (21) we get:
\bea
\left\{|\alpha(\vk,\tau)|^{2}\atop|\beta(\vk,\tau)|^{2}\right\}=
\frac{\sinh\left(\frac{\pi\tau}{4}(\pm E_{f}+E{i}-m_{f}+m_{i})\right)
      \sinh\left(\frac{\pi\tau}{4}(E_{f}\pm E{i}\pm m_{f}\mp m_{i})\right)}
     {\sinh\left(\frac{\pi\tau}{2} E_{f}\right)
      \sinh\left(\frac{\pi\tau}{2} E_{i}\right)}
\label{eq:23}
\eea
The Bogoliubov coefficients in (23) satisfy the crucial unitarity condition (10).
In the limit of step-like transition they correspond to (9) as one can check
by straightforward calculation.

As can be seen from (23), the coefficient $\beta(\vk,\tau)\rightarrow 0$
at $\vk\rightarrow 0$ and it falls down for large $\vk$:
\be
|\beta(\vk,\tau)|\approx r(\vk,\tau)\approx
\frac{\sinh\left(\pi\tau |m_{i}-m_{f}|/4\right)}
     {\sinh\left(\pi\tau k/2\right)},\;k\gg m_{i},m_{f}
\label{eq:24}
\ee
In the region $m_{f}\ll k\ll m_{i}$, where the coefficient $\beta(\vk)$ and the
evolution parameter $r(\vk)$ are maximal for step-like transition, we get:
\be
|\beta(\vk,\tau)|^{2}\approx \frac{1}{2}-\frac{1}{2}\left(\frac{m_{f}}{k}
+\frac{k}{m_{i}}h(\pi m_{i}\tau/2)\right)
\label{eq:25}
\ee
with
\be
h(x)=\frac{1}{2}(1+x\coth x),\quad h>1
\label{eq:26}
\ee
where $m_{i}\gg m_{f}, k\tau\ll 1$ was taken in accordance with typical
parameters $m_{i}\sim 350 MeV, m_{f}\sim 5 MeV,k\sim 40 MeV$ in Section 2
and with typical time duration $\tau\sim 1 fm$. As a result the maximum
of $\beta(\vk,\tau)$ and $r(\vk,\tau)$ is reduced for finite $\tau$
and shifted to smaller momentum,
\be
|\beta_{max}(\vk,\tau)|^{2} \approx \frac{1}{2}-\sqrt{hm_{f}/m_{i}},\quad 
k_{m}\approx\sqrt{m_{i}m_{f}/h}
\label{eq:27}
\ee
due to $h$-factor. At large $k\gg\tau^{-1}$ the effect is exponentially small,
$|\beta|\sim exp(-\pi k\tau/2)$ for smooth transition.

\section{Fermion creation and their correlations}

Using Bogoliubov transformation (17) one can find the final-state fermion
momentum distributions and the final-state fermion correlations. We confine
ourselves to symmetric case when the fermions with opposite momenta are produced
in an equivalent way (say central collisions of identical nuclei). For simplicity
the Bogoliubov coefficients will be taken to be real-valued and $k=|\vk|$
dependent. We also consider the simple model -- fast simultaneous transition
of large homogeneous system at rest. The movement of the system can be taken
into account by shifting of each moving element to its rest frame and
integrating over proper times and space-time raidities of the elements of
the system as it was done for photon production~\cite{A2} in heavy ion
collisions.

The final-state momentum distribution of the fermions (single-particle
inclusive cross-sections) is given by (below we use notations $b_{\nu},d_{\nu}$
for operators having definite helicity $\nu=\pm 1/2$)
\be
N_{f,\nu}(\vk)=\frac{1}{\sigma}\frac{d\sigma_{\nu}}{d^{3}k}=
\langle b^{\dag}_{f,\nu}(\vk)b_{f,\nu}(\vk)\rangle
\label{eq:28}
\ee
where brackets mean averaging over initial state.
Using (8) we get:
\be
N_{f,\nu}(\vk)=|\alpha(\vk)|^{2}\langle b^{\dag}_{i,\nu}(\vk)b_{i,\nu}(\vk)\rangle
-|\beta(\vk)|^{2}\langle d^{\dag}_{i,\nu}(-\vk)d_{i,\nu}(-\vk)\rangle
+|\beta(\vk)|^{2}\frac{V}{(2\pi)^{3}}
\label{eq:29}
\ee
and similar expression for antifermions (with interchange $b\leftrightarrow d$).
The last term in {\em rhs\/} of (29) reflects the result of the vacuum rearrangement
(the initial vacuum of fermions having mass $m_{i}$ is not the ground state
of $b_{f},d_{f}$-operators). The factor $V/{(2\pi)^{3}}$ replaces
$\delta^{3}(0)$ in the case of large but finite volume $V$. 

It is convenient to introduce the level population function $n_{\nu}(\vk)$:
\be
N_{\nu}(\vk)=\frac{V}{(2\pi)^{3}}n_{\nu}(\vk)
\label{eq:30}
\ee
Then (29) and corresponding equation for antiparticles take the simple form
(independently for every helicity)
\be
n_{f,\nu}(\vk)=|\alpha(\vk)|^{2}n_{i,\nu}(\vk)+
               |\beta(\vk)|^{2}\left( 1-\bar{n}_{i,\nu}(-\vk)\right),
\label{eq:31}
\ee
\be
\bar{n}_{f,\nu}(\vk)=|\alpha(\vk)|^{2}\bar{n}_{i,\nu}(\vk)+
               |\beta(\vk)|^{2}\left( 1- n_{i,\nu}(-\vk)\right)
\label{eq:32}
\ee
where the notation $\bar{n}$ is used for antifermions. In general one has
to consider simultaneously the fermions and antifermions with momenta
$\pm \vk$ to get the full picture of the particle creation. For example,
if there are no particles in the initial state then we have $|\beta(\vk)|^{2}$
particles of each kind in the final state. If there is one particle having momentum
$\vk$ and helicity $\nu$ in the initial state then we have
\be
n_{f,\nu}(\vk)=1,\quad \bar{n}_{f,\nu}(-\vk)=0
\label{eq:33}
\ee
and $|\beta|^{2}$ particles in the rest of the states. If there is
a fermion-antifermion (singlet) pair having zero momentum in the initial state,
that is $n_{i,\nu}(\vk)=1, \; \bar{n}_{i,\nu}(-\vk)=1$, then we get
\be
n_{f,\nu}(\vk)= \bar{n}_{f,\nu}(-\vk)=|\alpha(\vk)|^{2},\quad
n_{f,\nu}(-\vk)= \bar{n}_{f,\nu}(\vk)=|\beta(\vk)|^{2}
\label{eq:34}
\ee
for the final state. Therefore in zero-momentum states which are initially
copletely occupied (both with fermions and antifermions) the occupation number
decreases contrary to initially empty states where the occupation number
increases. For termal equilibrium $n(\vk)$ is the Fermi distribution,
depending on temperature and chemical potential and the both effects are
present.

The transition effect is better seen in two-particle correlations.
Two-particle inclusive cross-sections are given by
\bea
\frac{1}{\sigma}\frac{d^{2}\sigma^{++}_{\mu\nu}}{d^{3}k_{1}d^{3}k_{2}}=
\langle b^{\dag}_{f,\nu}(\vkt)b^{\dag}_{f,\mu}(\vko)
b_{f,\mu}(\vko)b_{f,\nu}(\vkt)\rangle=   \nonumber \\
\langle b^{\dag}_{f,\nu}(\vkt)b_{f,\nu}(\vkt)\rangle
\langle b^{\dag}_{f,\mu}(\vko)b_{f,\mu}(\vko)\rangle -
\delta_{\mu\nu}\langle b^{\dag}_{f,\nu}(\vkt)b_{f,\mu}(\vko)\rangle
               \langle b^{\dag}_{f,\mu}(\vko)b_{f,\nu}(\vkt)\rangle
\label{eq:35}
\eea
for two fermions with helicities $\mu,\nu$ (correlation of identical fermions)
and
\bea
\frac{1}{\sigma}\frac{d^{2}\sigma^{+-}_{\mu\nu}}{d^{3}k_{1}d^{3}k_{2}}=
\langle b^{\dag}_{f,\nu}(\vkt)d^{\dag}_{f,\mu}(\vko)
d_{f,\mu}(\vko)b_{f,\nu}(\vkt)\rangle=   \nonumber \\
\langle b^{\dag}_{f,\nu}(\vkt)b_{f,\nu}(\vkt)\rangle
\langle d^{\dag}_{f,\mu}(\vko)d_{f,\mu}(\vko)\rangle +
\delta_{\mu\nu}\langle b^{\dag}_{f,\nu}(\vkt)d^{\dag}_{f,\mu}(\vko)\rangle
\langle d_{f,\mu}(\vko)b_{f,\nu}(\vkt)\rangle
\label{eq:36}
\eea
for fermion-antifermion pair. The last term in {\em rhs} of (36) is essential
only in the presence of the time evolution of the fermion field. Here for
simplicity we do not take into account the dynamical correlations arising
due to Coulomb and strong interactions of the fermions.

Considering correlations in finite volume $V$ we can use the modified
creation and annihilation operators~\cite{{AW},{A3}} which are nonzero
only in the volume $V$ and satisfy modified anticommutation relations
of the form
\be
\left[ b_{\mu}(\vko),b^{\dag}_{\nu}(\vkt)\right]_{+}=
\delta_{\mu\nu}\frac{V}{(2\pi)^{3}}G(\vko-\vkt), \quad G(0)=1
\label{eq:37}
\ee
where $G(\vk)$ is the normalized form-factor of the fermion source
(normalized Fourier transform of the source density). The corresponding
correlator of the modified creation and annihilation operators has the form
\be
\langle b^{\dag}_{\mu}(\vko),b_{\nu}(\vkt)\rangle \approx
\delta_{\mu\nu}\frac{V}{(2\pi)^{3}}n((\vko+\vkt)/2) G(\vko-\vkt)
\label{eq:38}
\ee
The right hand side of (37),(38) represents the smeared $\delta$-function
having the width of the order of the inverse size of the source which is much
less than the width of the momentum distribution $n(\vk)$.

Using Bogoliubov transformation (8), substituting (38) for arising fermion
correlators of initial-state operators and taking into account the small
width of the form-factor $G(\vk)$ one can express the final-state correlators
through Bogoliubov coefficients $\alpha, \beta$ and the form-factor $G$:
\bea
\langle b^{\dag}_{f,\nu}(\vkt)b_{f,\nu}(\vko)\rangle
\langle b^{\dag}_{f,\nu}(\vko)b_{f,\nu}(\vkt)\rangle \nonumber \\
=\frac{V^{2}}{(2\pi)^{6}}\left(|\alpha(\vk)|^{2}n_{i,\nu}(\vk)+
|\beta(\vk)|^{2}(1-\bar{n}_{i,\nu}(\vk))\right)^{2}|G(\vko-\vkt)|^{2},
\label{eq:39}
\eea
\bea
\langle b^{\dag}_{f,\nu}(\vkt)d^{\dag}_{f,\nu}(\vko)\rangle
\langle d_{f,\nu}(\vko)b_{f,\nu}(\vkt)\rangle \nonumber \\
=\frac{V^{2}}{(2\pi)^{6}}|\alpha(\vk)\beta(\vk)(1-n_{i,\nu}(\vk)-
 \bar{n}_{i,\nu}(\vk))G(\vko+\vkt)|^{2}
\label{eq:40}
\eea
where smooth functions $\alpha(\vk),\beta(\vk)$ can be evaluated at any of momenta
$\vko,\vkt\approx\pm\vk$ and we suggest that the process is $\vk\leftrightarrow
-\vk$ symmetric. Taking sum over helicities $\nu$ we finally obtain the relative
(divided by the product of single-particle distributions) correlation functions
which are measured in experiment:
\be
C^{++}(\vko,\vkt)=1-\frac{1}{2}|G(\vko-\vkt)|^{2},
\label{eq:41}
\ee
\be
C^{+-}(\vko,\vkt)=1+\frac{1}{2}R^{2}(\vk)|G(\vko+\vkt)|^{2}
\label{eq:42}
\ee
with
\be
R^{2}(\vk)=\frac{|\alpha(\vk)\beta(\vk)(1-n_{i}(\vk)-\bar{n}_{i}(\vk))|^{2}}
{\left[|\alpha(\vk)|^{2}n_{i}(\vk)+|\beta(\vk)|^{2}(1-\bar{n}_{i}(\vk))\right]
 \left[|\alpha(\vk)|^{2}\bar{n}_{i}(\vk)+|\beta(\vk)|^{2}(1-n_{i}(\vk))\right]}
\label{eq:43}
\ee
where we suggest that the helicities $\pm1/2$ are equally represented in the
initial state.

The equation (41) represents the identical particle correlation in
its simplest (no interaction) form. The minus sign is a distinctive feature
of the fermions and the factor $1/2$ arises due to sum over polarizations
(only two of four final polarization states contain identical particles).

The equation (42) describes the time evolution effect which depends
on evolution parameter $r$ (or $\beta,\alpha$). For the vacuum initial state
the relative correlation function is big being equal to $\alpha^{2}/2\beta^{2}>
1/2$. It is especially big for small $\beta$ but in this case the pair
production itself is weak, see (31),(32). In the case of cold dense matter the
correlation function is big for empty (high momentum) initial states.
In the case of hot quark-gluon plasma in the initial state one has to take
into account the presence of effective termal quark mass which is much
bigger than original quark mass ($m_{i}\sim gT$ for light quarks where
$g$ is QCD coupling constant and $T$ is the temperature).

\section{Conclusion}
Calculation of the evolution effect for fermions shows that opposite-side
fermion-antifermion correlations can be large.
They can serve as a sign of chiral phase transition in quantum chromodynamics.

The work was supported by the Russian Fund for Basic Research (grant 02-02-16779).


\begin{thebibliography}{99}
%\bibitem{CMM}
% L.P.Csernai, I.N.Mishustin, and A.Mocsy, in {\it Proceedings of the 4th
% Rio de Janeiro Int. Workshop on Relativistic Aspects of Nuclear Physics},
% (World Sci., Singapore, 1995), p.137.
\bibitem{AW}
 I.V.Andreev and R.M.Weiner,  Phys. Lett. B {\bf 373}, 159 (1996).
\bibitem{AC}
 M.Asakawa and T.Csorgo,  Heavy Ion Phys. {\bf 4}, 233 (1996).
\bibitem{A1}
 I.V.Andreev,  Mod. Phys. Lett. A {\bf 14}, 459 (1999).
\bibitem{ACG}
 M.Asakawa, T.Csorgo, and M.Gyulassy, Phys. Rev. Lett. {\bf83}, 4019 (1999).
\bibitem{A2}
 I.V.Andreev, hep-ph/0105208, to be published in Phys. At. Nucl. (2002)
\bibitem{A3}
 I.V.Andreev,  Phys. At. Nucl. {\bf 63}, 1988 (2000).
\bibitem{B}
 N.N.Bogoliubov, Proc. Acad. Sci. USSR, Ser. Phys. {\bf 11}, 77 (1947).
\bibitem{N}
A.I.Nikishov, in {\it Proceedings of the P.N.Lebedev Physical Institute},
 (Nauka, Moscow, 1979), vol.111, p.152.
%\bibitem{LL}
% L.D.Landau, E.M.Lifshits, {\it Quantum mechanics}
% (Nauka, Moscow, 1974), p.746.

\end{thebibliography}
\end{document}